\documentclass{IEEEtran}
\usepackage{amssymb}
\usepackage{}
\usepackage{amsfonts}
\usepackage{amsmath}
\usepackage{algorithm}
\usepackage{algorithmic}
\usepackage{cite}

\usepackage{graphicx,subfigure,amsmath,amssymb,cite}

\usepackage[dvips]{color}
\usepackage{float}
\ifCLASSINFOpdf
\else
\fi
\begin{document}

\title{Power Allocation Strategies for Secure Spatial Modulation}

\author{Guiyang Xia, Linqiong Jia, Yuwen Qian, Feng Shu, Zhihong Zhuang, Jiangzhou Wang, \emph{Fellow}, \emph{IEEE}

\thanks{Guiyang Xia, Linqiong Jia, Yuwen Qian, Feng Shu and Zhihong Zhuang are with the School of Electronic and Optical Engineering, Nanjing University of Science and Technology, 210094, CHINA. Feng Shu is also with the School of Computer and  information at Fujian Agriculture and Forestry University, Fuzhou, 350002,  China. 
}
\thanks{Jiangzhou Wang is with the School of Engineering and Digital Arts, University of Kent, Canterbury CT2 7NT, U.K. 
}
}

\markboth{}%
{Shell \MakeLowercase{\textit{et al.}}: Bare Demo of IEEEtran.cls for Journals}

\maketitle

\begin{abstract}
In secure spatial modulation (SM) networks, power allocation (PA) strategies are investigated in this paper under the total power constraint. Considering that there is no closed-form expression for secrecy rate (SR), an approximate closed-form expression of SR is presented, which is used as an efficient metric to optimize PA factor and can greatly reduce the computation complexity. Based on this expression, a convex optimization (CO) method of maximizing SR (Max-SR) is proposed accordingly. Furthermore, a method of maximizing the product of  signal-to-leakage and noise ratio (SLNR) and artificial noise-to-leakage-and noise ratio (ANLNR) (Max-P-SAN) is proposed to provide an analytic solution to PA with extremely low-complexity. Simulation results demonstrate that the SR performance of the proposed CO method is close to that of the optimal  PA strategy of  Max-SR with exhaustive search  and   better than that of Max-P-SAN in the high signal-to-noise ratio (SNR) region. However, in the low and medium SNR regions, the SR performance of the proposed Max-P-SAN  slightly exceeds that of the proposed CO.
\end{abstract}


\IEEEpeerreviewmaketitle

\section{Introduction}

\IEEEPARstart{A}{s} a promising and green technology in multiple-input-multiple-out (MIMO) systems, spatial modulation (SM) \cite{Mesleh2008Spatial} exploits both the index of activated antenna and amplitude phase modulation (APM) symbol to transmit messages. Due to the broadcasting characteristic of wireless transmission \cite{Wang2012Distributed, QQWu2017, Aghdam2017Joint} , physical layer security becomes an urgent and important problem in wireless communication \cite{ChenXiaoming, SHYan2016, YongpengWu2017}.

How to make SM have a capability to achieve a secure transmission become an important issue for SM networks. In \cite{Wang2015Secrecy}, without the knowledge of Eve's location, the confidential messages are securely transmitted from the SM transmitter to the desired receiver by projecting artificial noise (AN) onto the null-space of the desired channel. In \cite{Liu2017Secure}, the authors proposed a full-duplex desired receiver, where the confidential messages is received and meanwhile AN is emitted to corrupt the illegal receiver (Eve). This scheme can provide a high capability to combat eavesdropping.  The authors in \cite{Wangzw2018} proposed two new schemes of transmit antenna selection for secure SM networks: maximizing secrecy rate (SR) and leakage, where the proposed leakage-based antenna selection scheme achieve an excellent SR performance with a very low-complexity.

As an efficient way to improve the security of SM networks, power allocation (PA) has an important impact on SR performance. However, there are only little research of making an investigation on PA strategies for SM. In \cite{Wu2015Secret}, the optimal PA factor was given for precoding SM by maximizing the SR performance with exhaustive search (ES). Thus,  no closed-form SR expression can be developed for PA, which will result in a high computational complexity to complete ES. This motivates us to find some closed-form solutions or low-complexity iterative methods for different PA strategies. In this paper, we will focus on the investigation of PA strategies in secure SM (SSM) networks. The main contributions are summarized as follows:
\begin{enumerate}
\item Due to the fact that SR lacks a closed-form expression in SSM systems, its effective approximate simple expression is defined as a metric, which can dramatically reduce the evaluation complexity of SR values to optimize PA factor. Following this definition, a convex optimization (CO) method is proposed to address the optimization problem of maximizing SR (Max-SR). The simulation results show that the SR difference between the proposed CO and the Max-SR with ES can be negligible for almost all SNR regions.
\item To reduce the computational complexity of the above CO method and at the same time provide a closed-form PA strategy, a PA strategy of maximizing the product of signal-to-leakage plus noise ratio (SLNR) and AN-to-leakage plus noise ratio (ANLNR) (Max-P-SAN) is proposed, which can strike a good balance between maximizing SLNR and maximizing ANLNR. Its analytic expression of PA factor is also given. Simulation results show that the SR performance of Max-P-SAN method, with extremely low complexity, tends to that of Max-SR with ES method and is slightly better than that of CO in the low and medium SNR regions.
\end{enumerate}

The reminder is organized as follows. In Section II, a SSM system with the aid of AN is described. In Section III, first, the approximate simple formula of average SR is given, and two PA strategies, CO and Max-P-SAN, are proposed to maximize approximate SR and the product of SLNR and ANLNR, respectively. Subsequently, numerical simulations and analysis are presented in Section IV. Finally, we make our  conclusions in Section V.

\section{System Model}

\begin{figure} [ht]
 \centering
 \includegraphics[width=0.48\textwidth]{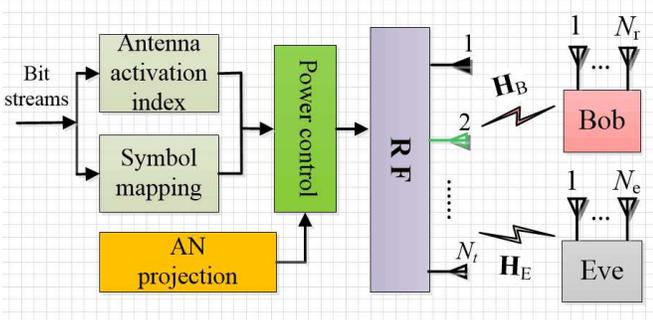}\\
 \caption{A secure SM system model.}\label{SystemModel}
\end{figure}

Fig.~\ref{SystemModel} sketches a SSM system with $N_t$  transmit antennas (TAs) at Alice, $N_r$  receive antennas (RAs) at Bob, and $N_e$  RAs at Eve, respectively. Here, Eve intends to intercept the confidential messages. Additionally, we denote the size of signal constellation $\mathcal{M}$ by $M$. As a result, $\textrm{log}_2N_tM$ bits per channel use can be transmitted, where $\textrm{log}_2N_t$ bits are used to select one active antenna, and the remaining $\textrm{log}_2M$ bits are used to form a constellation symbol.

Referring to the SSM model in \cite{Wangzw2018}, the transmit signal with the aid of AN is represented by
\begin{align} \label{x}
\textbf{s}=\sqrt{\beta P}\textbf{e}_ib_j+ \sqrt{(1-\beta)P}\textbf{T}\textbf{n}
\end{align}
where $\beta\in [0, 1]$ is the PA factor, $P$ denotes the total transmit power. $\textbf{e}_i$ is the $i$th column of identity matrix $\textbf{I}_{N_t}$, which means the $i$th antenna is chosen for transmitting symbol $b_j$, which is the input symbol equiprobably drawn from a $M$-ary constellation. $\textbf{T}$ is the projection matrix of the AN vector $\textbf{n} \in \mathbb{C}^{N_t \times 1}$ with $\textrm{tr}(\textbf{T}\textbf{T}^H)=1$, where $\textrm{tr}(\cdot)$ denotes the matrix trace. The  receive vector of symbols at the desired and eavesdropping receivers are
\begin{align} \label{y_B}
&\textbf{y}_B=\sqrt{\beta P}\textbf{H}_B\textbf{e}_ib_j+\sqrt{(1-\beta)P}\textbf{H}_B\textbf{T}\textbf{n}+\textbf{n}_B
\\ \label{y_E}
&\textbf{y}_E=\sqrt{\beta P}\textbf{H}_E\textbf{e}_ib_j+ \sqrt{(1-\beta)P}\textbf{H}_E\textbf{T}\textbf{n}+ \textbf{n}_E
\end{align}
where $\textbf{H}_B \in \mathbb{C}^{N_r \times N_t}$ and $\textbf{H}_E \in \mathbb{C}^{N_e \times N_t}$ are the complex channel gain matrices from Alice to Bob and from Alice to Eve, with each elements of $\textbf{H}_B$ and $\textbf{H}_E$ obeying the Gaussian distributions with zero mean and unit variance, i.e., $\mathcal{CN}(0,1)$. Additionally, $\textbf{n}_B \in \mathbb{C} ^{N_r \times 1}$, and  $\textbf{n}_E \in \mathbb{C} ^{N_e \times 1}$ are complex Gaussian noise at desired and eavesdropping receivers with $\textbf{n}_B \sim \mathcal{CN}(0,\sigma_B^2\textbf{I}_{N_r})$ and $\textbf{n}_E \sim \mathcal{CN}(0,\sigma_E^2\textbf{I}_{N_e})$, respectively.
Given a specific channel realization,  the mutual information between Alice and  Bob, and between Alice and Eve are
\begin{align} \label{I_B} \nonumber
I_B&(\textbf{s};\textbf{y}_B'|\textbf{H}_B)=\textrm{log}_2N_tM - \\ &  \frac{1}{N_tM}\sum \limits_{i = 1}^{N_tM} {\mathbb{E}_{\textbf{n}_B'} \left\{ \textrm{log}_2 \sum \limits_{j=1} ^{N_tM} \textrm{exp} \left( -f_{b,i,j}+\|\textbf{n}_B'\|^2 \right) \right\}}
\\ \label{I_E} \nonumber
I_E&(\textbf{s};\textbf{y}_E'|\textbf{H}_E)=\textrm{log}_2N_tM - \\ & \frac{1}{N_tM} \sum \limits_{m  = 1}^{N_tM} {\mathbb{E}_{\textbf{n}_E'} \left\{ \textrm{log}_2 \sum \limits_{k=1} ^{N_tM} \textrm{exp} \left( -f_{e,m,k}+\|\textbf{n}_E'\|^2 \right) \right\}}
\end{align}
where $f_{b,i,j}=\|\sqrt{\beta P}\textbf{W}_B^{-1/2}\textbf{H}_B\textbf{d}_{ij}+\textbf{n}_B'\|^2$, $f_{e,m,k}=\|\sqrt{\beta P}\textbf{W}_E^{-1/2}\textbf{H}_E
\textbf{d}_{mk}+\textbf{n}_E'\|^2$, $\textbf{d}_{ij}=\textbf{x}_i-
\textbf{x}_j$, and $\textbf{d}_{mk}=\textbf{x}_m-
\textbf{x}_k$.  Here, $\textbf{x}_i$,~$\textbf{x}_j$,~$\textbf{x}_m$, or ~$\textbf{x}_k$ is one possible transmit vectors in the set of combining antenna and all possible symbol vectors. $\textbf{W}_B$ is the covariance matrix of the last two terms of $\textbf{y}_B $ in (2), i.e. AN plus noise, and $\textbf{W}_B=(1-\beta)P\textbf{C}_B + \sigma_B^2\textbf{I}_{N_r}$,
where
$\textbf{C}_B=\textbf{H}_B\textbf{T}\textbf{T}^H\textbf{H}_B^H$. Similarly, $\textbf{W}_E=(1-\beta)P\textbf{C}_E + \sigma_E^2\textbf{I}_{N_e}$ where $\textbf{C}_E=\textbf{H}_E\textbf{T}\textbf{T}^H\textbf{H}_E^H$.
 From \cite{Wang2015Secrecy}, it is known that pre-multiplying $\boldsymbol{y}_B $ in (2) by  $\textbf{W}_B^{-1/2}$ from left is to whiten the AN plus noise into a white Gaussian noise. The linear transformation does not change the mutual information, thus $I(\textbf{x};\textbf{y}_B)=I(\textbf{x};\textbf{y}_B')$, where $\textbf{y}_B'=\textbf{W}_B^{-1/2}\textbf{y}_B$, and
 $\textbf{n}_B'=\textbf{W}_B^{-1/2}(\sqrt{(1-\beta)P} \textbf{H}_B\textbf{T}\textbf{n}+ \textbf{n}_B )$.
Similarly, $I(\textbf{x};\textbf{y}_E)=I(\textbf{x};\textbf{y}_E')$. The average SR is defined as
\begin{align}\label{SR}
\bar{R}_s=\mathbb{E}_{\textbf{H}_B,\textbf{H}_E}\left[ I(\textbf{x};\textbf{y}_B)- I(\textbf{x};\textbf{y}_E),0 \right]^+.
\end{align}
where $\left[ a \right]^+$=max(a,0) and $R_s(\beta) =I(\textbf{x};\textbf{y}_B)- I(\textbf{x};\textbf{y}_E)$ is the instantaneous SR for a specific channel realization. Here, we assume that the ideal channel knowledge of $\textbf{H}_B$ and $\textbf{H}_E$ per channel use are available at transmitter \cite{YongpengWu2017}. In accordance with the above equations, the optimization problem of maximizing SR over PA factor can be casted as
\begin{align}
 \max \limits_\beta ~  R_s(\beta)~~~~~\textrm{s.t.} ~~ 0 \leq \beta \leq 1.   \label{P0}
\end{align}

\section{Power allocation strategy for secrecy rate maximization}

In this section, two new PA methods, called CO and Max-P-SAN, are proposed. The former forms an iterative solution, and the latter produces a closed-form PA expression.

\subsection{Proposed CO method}
Due to the absence of closed-form expression of SR, it is hard for us to design an efficient method to optimize PA factor directly. Although the ES method in \cite{Wu2015Secret} is employed to find out the optimal PA factor for a given SNR, but its high complexity limits its applications to practical SM systems. In view of this, the cut-off rate \cite{Aghdam2017Joint} with closed-form for traditional MIMO systems can be easily extended to the secure SM systems, and may be adopted as an efficient metric to optimize the PA factor  as follows
\begin{align} \label{R_s_appro}
R_s^{a}(\beta)=I_0^{B}-I_0^{E},
\end{align}
where $I_0^{B}$ is the cut-off rate for the desired receiver given by
\begin{align}
I_0^{B}=\zeta- \textrm{log}_2\sum \limits_{i=1}^{N_tM}\sum \limits_{j=1}^{N_tM}\textrm{exp}\left( \frac{-\beta P}{4}\textbf{d}_{ij}^H\textbf{H}_B^H \boldsymbol{\omega}_B \textbf{H}_B\textbf{d}_{ij} \right)
\end{align}
where $\zeta=2\textrm{log}_2N_tM$, $\boldsymbol{\omega}_B(\beta)=\textbf{W}_B^{-1}$. Similarly, the cut-off rate $I_0^E$ for the eavesdropper is
\begin{align}
I_0^{E}=\zeta- \textrm{log}_2\sum \limits_{m=1}^{N_tM}\sum \limits_{k=1}^{N_tM}\textrm{exp}\left( \frac{-\beta P}{4}\textbf{d}_{mk}^H\textbf{H}_E^H \boldsymbol{\omega}_E \textbf{H}_E\textbf{d}_{mk} \right)
\end{align}
where $\boldsymbol{\omega}(\beta)=\textbf{W}_E^{-1}$. The detailed process of (\ref{R_s_appro}) refers to Appendix A in \cite{Aghdam2017Joint}. Replacing the objective function in (\ref{P0}) by (\ref{R_s_appro}) yields
\begin{align}\label{P1}
\max \limits_\beta ~ R_s^{a}(\beta) ~~~~\textrm{s.t.}~~  0\leq \beta \leq1.
\end{align}
However, the objective function of problem (\ref{P1}) is non-concave. Note that $\boldsymbol{\omega}_E(\beta) \approx \frac{1}{(1-\beta)P}\textbf{C}_E^{-1}$ in the high SNR region (i.e, $\sigma_B^2 \rightarrow 0$) when $\textbf{C}_E$ is nonsingular, thus we have
\begin{align}  \label{Equ_E} \nonumber
\tilde{\kappa}_E&(\beta) \approx\textrm{log}_2 \kappa_E \\
& =\textrm{log}_2\sum \limits_{m=1}^{N_tM}\sum \limits_{k=1}^{N_tM}\textrm{exp}\left( \frac{-\beta \textbf{d}_{mk}^H\textbf{H}_E^H\textbf{C}_E^{-1}\textbf{H}_E\textbf{d}_{mk}}{4(1-\beta)} \right).
\end{align}
It can be seen that $\tilde{\kappa}_E(\beta)$ is convex with respect to $\beta$ and then the objective function becomes a difference between two convex functions. To convert this difference to a concave function, we have the linear under-estimator of $\tilde{\kappa}_E(\beta)$ at the feasible point $\beta_{k-1}$ as follows
\begin{align}  \label{Linear_E}
\tilde{\kappa}_E \geq \tilde{\kappa}_E(\beta_{k-1}) + \tilde{\kappa}'_E (\beta_{k-1}) (\beta_k-\beta_{k-1})=g_E(\beta_k)
\end{align}
where $\tilde{\kappa}'_E (\beta_{k-1})$ is the first derivative value of function $\tilde{\kappa}_E$ at $\beta_{k-1}$, and
\begin{align}
\kappa_E'=\frac{1}{\textrm{ln}2\cdot\kappa_E}\sum \limits_{m=1}^{N_tM}\sum \limits_{k=1}^{N_tM} \frac{-PQ_{mk}}{4(1-\beta)^2} \textrm{exp} \left( \frac{-\beta P Q_{mk} }{4(1-\beta)} \right)
\end{align}
where $Q_{mk}=\textbf{d}_{mk}^H\textbf{H}_E^H\textbf{C}_E^{-1}\textbf{H}_E\textbf{d}_{mk}$. For a given feasible solution $\beta_0$, the problem (\ref{P1}) can be solved by the following approximate iterative sequence of convex problems
\begin{align}
\max \limits_\beta ~ G(\beta_k) = g_E(\beta_k) - \tilde{\kappa}_B(\beta_k)
~~~\textrm{s.t.} ~~ 0< \beta_k <1. \label{P3}
\end{align}
It is clear that the objective function in (\ref{P3}) is concave. Starting with an feasible point $\beta_0$, the optimization problem (\ref{P3}) is iteratively solved with different $\beta_k$, where $\left\{\beta_k\right\}$ is the generated sequence of solutions corresponding to the $k$th iteration. This iterative process  terminates until $
\mid G(\beta_{k})-G(\beta_{k-1}) \mid  \leq \varepsilon $, where $\varepsilon$ is a prechosen threshold.

\subsection{Proposed Max-P-SAN method}

Utilizing the leakage idea \cite{Sadek2007, ShuPower2011}, the SLNR, mainly denoting the desired signal leakage to the eavesdropping direction, is given by
\begin{align}
\textrm{SLNR}_B= \textrm{tr} (\textbf{H}_B^H\textbf{H}_B) \left( \textrm{tr} (\textbf{H}_E^H\textbf{H}_E)+ \sigma_B^2N_tN_r/ \beta P  \right)^{-1}
\end{align}
In the same manner, the AN is viewed as the useful signal of the eavesdropper, the ANLNR from the wiretap channel to the desired channels is as follows
\begin{align}
\textrm{ANLNR}_E= \textrm{tr} (\textbf{C}_E) \left( \textrm{tr} (\textbf{C}_B)+\sigma_E^2N_e /(1-\beta) P \right)^{-1}
\end{align}
It is hard to jointly optimize the two objective functions $\textrm{SLNR}_B$ and $\textrm{ANLNR}_E$. To simplify the joint optimization problem, we  multiply the two functions to form a new product of $\textrm{SLNR}_B$ and $\textrm{ANLNR}_E$, which is used as a single objective function. This will significantly simplify our optimization manipulation. Maximizing their product means maximizing at least one of them, or both them. From simulations, we find that the proposed product method performs very well, and make a good balance between performance and complexity.  Then, the associated optimization problem can be written as
\begin{align} \nonumber
 \max \limits_\beta ~~ & F(\beta) =\frac{\kappa_B \cdot \beta}{\kappa_E \cdot \beta
+\sigma_B^2N_tN_r}  \cdot
\frac{(1-\beta)\cdot \omega_E}{(1-\beta)\cdot \omega_B
+\sigma_E^2N_e} \\
& \textrm{s.t.} ~~ 0\leq \beta \leq1.  \label{P_Leakage}
\end{align}
where
$\kappa_B=\frac{P}{N_t}\textrm{tr}(\textbf{H}_B^H\textbf{H}_B)$, $\kappa_E=\frac{P}{N_t}\textrm{tr}(\textbf{H}_E^H\textbf{H}_E)$,
$\omega_B=P\textrm{tr}(\textbf{C}_B)$, and
$\omega_E=P\textrm{tr}(\textbf{C}_E)$.
Therefore, the promising optimal values of $\beta$ in (\ref{P_Leakage}) should satisfy the following equation
\begin{align}  \label{Frist_D}
 F'(\beta)= \frac{\varphi_a\left( \varphi_o\beta^2-2\varphi_d \beta +\varphi_d \right)}{(-\varphi_b\beta^2+\varphi_c \beta+ \varphi_d)^2}=0
\end{align}
where $\varphi_a=\kappa_B\omega_E$, $\varphi_b=\kappa_E\omega_B$, $\varphi_c=\kappa_E\omega_B +\sigma_B^2\kappa_EN_e$, $\varphi_d=\sigma_B^2 \omega_B N_r + \sigma_B^2 \sigma_E^2 N_rN_e$, and $\varphi_o=\varphi_b-\varphi_c$. Based on  (\ref{Frist_D}), it is seen that the denominator of the derivative  and $\varphi_a$ in (\ref{Frist_D}) are both greater than 0, we only need to solve the roots of equation $\varphi_o\beta^2-2\varphi_d \beta +\varphi_d=0$. Due to $\varphi_o < 0$ and $\bigtriangleup=\varphi_d^2-\varphi_o\varphi_d > 0$, this equation has two real-valued roots. In summary,  the set of feasible solutions to (\ref{P_Leakage}) is
\begin{align} \nonumber
S=\left\{\beta_1=\frac{\varphi_d + \sqrt{\bigtriangleup}}{\varphi_o}, \beta_2=\frac{\varphi_d - \sqrt{\bigtriangleup}}{\varphi_o}, \beta_3=0, \beta_4=1 \right\},
\end{align}
where $\beta_1$, and $\beta_2$ are two solutions to the quadratic equation in (\ref{Frist_D}) while $\beta_3=0$, and $\beta_4=1$ are two end-points of the feasible search interval $[0, 1]$. Obviously, $\beta_3=0$ means that there is no confidential messages to be sent. In other words, SR=0.  Thus, this point can be directly removed from the solution set. $\beta_1<0$ falls outside the feasible set $[0, 1]$, and can be deleted directly. Considering the function $F(\beta)$ is a continuous and differentiable function over the interval $[0, 1]$, its first derivative is negative as $\beta$ goes to one from the left. Thus, $F(1)$  is the local minimum point, which rules out it from the the feasible solution set referring to the set of maximizing $F(\beta)$. Finally, we have the unique solution
\begin{align}
\beta_2=(\varphi_d - \sqrt{\bigtriangleup})/\varphi_o
\end{align}
due to the fact that $F'(\beta_2)=0$ and $\beta_2 \in [0, 1]$.


\subsection{Complexity Analysis and Comparison}
Below, we present a complexity comparison among the three methods: CO, Max-P-SAN, and ES. Firstly, the complexity of the ES method in \cite{Wu2015Secret} is about $\mathcal{C}_{\textrm{ES}}=2N_t^2M^2 l N_{\textrm{samp}}$ $[ 2(N_r+N_e)N_t^2+N_r+N_e]$ floating-point operations (FLOPs), where $l$ denotes the number of searches  depending on the required accuracy,  and $N_{\textrm{samp}} (\geq 500)$ is the number of realizations of noise sample points for accurately estimating expectation operators. For the proposed CO, its computational complexity is approximated as $\mathcal{C}_{\textrm{CO}}=3N_t^2M^2D_{\textrm{ite}}(2N_t^2+2N_t)$ FLOPs, where $D_{\textrm{ite}}$ is the number of iterations. Finally, it is obvious that the proposed Max-P-SAN scheme has the lowest complexity among the three methods, and its complexity is $\mathcal{C}_{\textrm{Max-P-SAN}}=2N_t^2(2N_r+3N_e)+2N_r^2N_t+2N_e^2N_t+N_t+N_r+N_e$ FLOPs. From the three complexity expressions,   the dominant term in $\mathcal{C}_{\textrm{Max-P-SAN}}$ is only quadratic. In general, $N_{\textrm{samp}} \gg N_t > N_r (N_e)$,  their complexities have an increasing order as follows: Max-P-SAN, CO, and ES.


\section{Simulation Results}
In what follows, numerical simulations are presented to evaluate the SR performance for two proposed PA strategies, with ES method as a  performance benchmark. Specially,  the noise variances are assumed to be identical, i.e., $\sigma_B^2=\sigma_E^2$. Modulation type is quadrature phase shift keying (QPSK).
\begin{figure} [ht]
 \centering
 \includegraphics[width=0.478\textwidth]{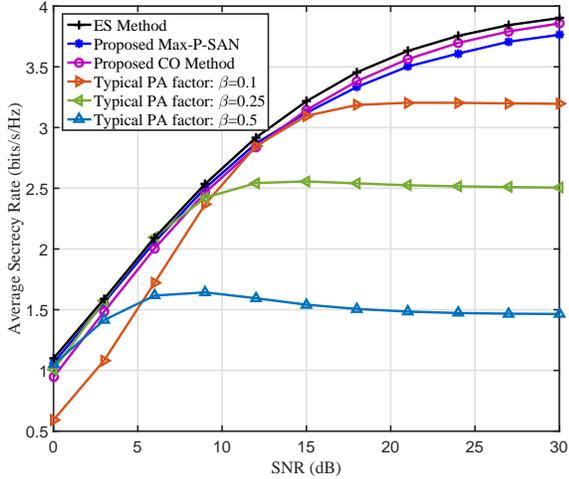}\\
 \caption{ Comparison of average SR for various PA methods with $N_t=4, N_r=2$ and $N_e=2$.}\label{4111}
\end{figure}

Fig. \ref{4111} plots the curves of SR versus SNR with $N_t=4, N_r=2$, and $N_e=2$. Here, three typical PA strategies, $\beta=0.1$, $0.25$, and $0.5$, are used as performance references.  From Fig. \ref{4111},  it is seen that the proposed CO method  can achieve the optimal SR performance being close to that of the ES method for almost all SNR regions. The proposed Max-P-SAN method approaches the ES performance in the low and medium SNR regions, but slightly worse than the ES in the high SNR region in terms of SR.  Because Max-P-SAN has a closed-form expression, it strikes a good balance between performance and complexity. Also, the two proposed methods perform much better than three typical fixed PA schemes. This means that they can harvest appreciable SR performance gains.

\begin{figure} [ht]
 \centering
 \includegraphics[width=0.478\textwidth]{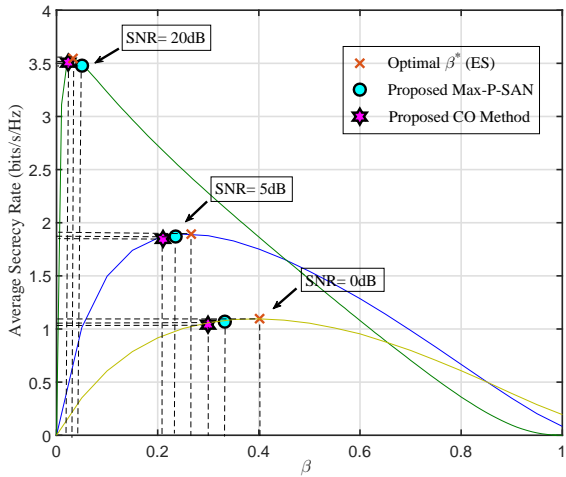}\\
 \caption{Comparison of the achievable SR for SNR=0, 5, 20dB with same configuration as Fig. \ref{4111}.}\label{CDF}
\end{figure}

Fig. \ref{CDF} plots the curves of maximum achievable SR versus $\beta$  for three given  SNRs: 0dB, 5dB, and 20dB, with ES as a performance benchmark. From this figure, it is obvious that all optimal values of $\beta$ reduce as SNR increases from 0dB to 20dB. This can be readily explained as follows: a high SNR means a good channel quality. This implies that less power is required to transmit confidential messages, and more power is utilized to emit AN to corrupt eavesdroppers.  Additionally, as SNR increases, the optimal values of $\beta$ corresponding to the two proposed PA are closer to that of $\beta$ for ES.


\section{Conclusion}
In this paper, we have made an investigation of PA strategies for SSM system. An efficient approximated expression of SR was given to simplify the computational complexity for optimizing PA factor. Then, two PA strategies were proposed to implement PA between confidential messages and AN. The first one is CO and the second one is Max-P-SAN. The former is iterative while the latter is closed-form. In accordance with simulations, we find: the proposed CO provides a SR performance being close to the ES method for almost all SNR regions, and the proposed Max-P-SAN can achieve the optimal SR performance in the low and medium regions with an extremely low-complexity. These strategies can be applied to the future SSM-based networks such as unmanned aerial vehicle, future mobile networks, and intelligent transportation.

\bibliographystyle{IEEEtran}
\bibliography{IEEEabrv,refer}

\ifCLASSOPTIONcaptionsoff
  \newpage
\fi

\end{document}